\documentclass[twocolumn]{aastex6}

\usepackage{natbib}
\usepackage{nicefrac}
\usepackage{amsmath}

\hypersetup{citecolor=blue,linkcolor=blue,urlcolor=blue}
\usepackage[hang]{footmisc}

\renewcommand{\footnotesize}{\scriptsize}
\renewcommand{\textsc}{}

\def\Msun{{\rm M}_\odot}

\shorttitle{{\sc Las Cumbres Observatory GW Followup}}
\shortauthors{{\sc Arcavi et al.}}

\defcitealias{G16}{G16}

\begin{document}

\title{Optical Follow-up of Gravitational-wave Events with Las Cumbres Observatory}

\author{
Iair~Arcavi\altaffilmark{1,2,12},
Curtis~McCully\altaffilmark{1,2},
Griffin~Hosseinzadeh\altaffilmark{1,2},
D.~Andrew~Howell\altaffilmark{1,2},
Sergiy~Vasylyev\altaffilmark{1,2},
Dovi~Poznanski\altaffilmark{3},
Michael~Zaltzman\altaffilmark{3},
Dan~Maoz\altaffilmark{3},
Leo~Singer\altaffilmark{4,5},
Stefano~Valenti\altaffilmark{6},
Daniel~Kasen\altaffilmark{7,8},
Jennifer~Barnes\altaffilmark{9,12},
Tsvi~Piran\altaffilmark{10}
and
Wen-fai~Fong\altaffilmark{11,13}
}

\affil{
\altaffilmark{1}{Department of Physics, University of California, Santa Barbara, CA 93106-9530, USA; \href{mailto:arcavi@ucsb.edu}{arcavi@ucsb.edu}}\\
\altaffilmark{2}{Las Cumbres Observatory, 6740 Cortona Drive, Suite 102, Goleta, CA 93117-5575, USA}\\
\altaffilmark{3}{Raymond and Beverly Sackler School of Physics and Astronomy, Tel Aviv University, Tel Aviv 69978, Israel}\\
\altaffilmark{4}{Joint Space-Science Institute, University of Maryland, College Park, MD 20742, USA}\\
\altaffilmark{5}{Astrophysics Science Division, NASA Goddard Space Flight Center, Code 661, Greenbelt, MD 20771, USA}\\
\altaffilmark{6}{Department of Physics, University of California, 1 Shields Avenue, Davis, CA 95616-5270, USA}\\
\altaffilmark{7}{Nuclear Science Division, Lawrence Berkeley National Laboratory, Berkeley, CA 94720-8169, USA}\\
\altaffilmark{8}{Departments of Physics and Astronomy, University of California, Berkeley, CA 94720-7300, USA}\\
\altaffilmark{9}{Columbia Astrophysics Laboratory, Columbia University, New York, NY, 10027, USA}\\
\altaffilmark{10}{Racah Institute of Physics, The Hebrew University of Jerusalem, Jerusalem 91904, Israel}\\
\altaffilmark{11}{CIERA and Department of Physics and Astronomy, Northwestern University 2145 Sheridan Road, Evanston, IL 60208, USA}\\
\altaffilmark{12}{Einstein Fellow}\\
\altaffilmark{13}{Hubble Fellow}\\
}

\begin{abstract}
We present an implementation of the \cite{G16} galaxy-targeted strategy for gravitational-wave (GW) follow-up using the Las Cumbres Observatory global network of telescopes. We use the Galaxy List for the Advanced Detector Era (GLADE) galaxy catalog, which we show is complete (with respect to a Schechter function) out to $\sim300$\,Mpc for galaxies brighter than the median Schechter function galaxy luminosity. We use a prioritization algorithm to select the galaxies with the highest chance of containing the counterpart given their luminosity, their position, and their distance relative to a GW localization, and in which we are most likely to detect a counterpart given its expected brightness compared to the limiting magnitude of our telescopes. This algorithm can be easily adapted to any expected transient parameters and telescopes. We implemented this strategy during the second Advanced Detector Observing Run (O2) and followed the black hole merger GW170814 and the neutron star merger GW170817. For the latter, we identified an optical kilonova/macronova counterpart thanks to our algorithm selecting the correct host galaxy fifth in its ranked list among 182 galaxies we identified in the Laser Interferometer Gravitational-wave Observatory LIGO-Virgo localization. This also allowed us to obtain some of the earliest observations of the first optical transient ever triggered by a GW detection (as presented in a companion paper).

\end{abstract}

\section{Introduction}

With the Advanced Laser Interferometer Gravitational-wave Observatory \citep[LIGO;][]{LIGO2015} providing detections of gravitational waves (GWs) since 2015 September \citep[e.g.][]{Abbott2016} and with Advanced Virgo \citep{Acernese2015} online since 2017 August, it is now feasible to search for electromagnetic (EM) counterparts to GW signals. The main sources of GWs detectable by advanced LIGO/Virgo are mergers of neutron stars and black holes. Of those, neutron star -- neutron star (NS--NS) and some neutron star -- black hole (NS--BH) mergers are expected to produce electromagnetic signatures. 

In both cases, emission is expected mainly from the radioactive decay of heavy elements formed through the $r$-process in the merger, as a small amount ($M\sim10^{-4}$--$10^{-2}\Msun$) of neutron-rich material is released at high velocities ($0.1$--$0.3c$) during the final coalescence \citep[e.g.][]{Rosswog1999, Rosswog2005, Hotokezaka2013, Sekiguchi2016}, and possibly also in outflows from an accretion disk \citep[e.g.][]{Metzger2008, Grossman2014, Kasen2015}. Following the decompression of the ejecta from nuclear densities, rapid neutron capture ($r$-process) leads to the formation of heavy radioactive elements which then release energy as they decay, powering an electromagnetic light curve \citep[e.g.][]{Li1998, Rosswog2005, Metzger2010, Goriely2011, Roberts2011, Metzger2012, Rosswog2013}. These events, which are predicted to be brighter than novae but fainter than supernovae, have been named ``macronovae'' \citep{Kulkarni2005} or ``kilonovae'' \citep{Metzger2010}. Additional emission sources such as free neutron decay leading to prompt blue emission \citep{Metzger2015} and magnetar spindown \citep{Metzger2014} have also been suggested. For recent reviews on kilonovae see \cite{Tanaka2016} and \cite{Metzger2017}.

The emission properties of a kilonova depend strongly on the composition of the elements produced in the merger, which is a major source of uncertainty in the models. Heavier elements known as lanthanides can increase the ejecta opacity by several orders of magnitude \citep{Kasen2013, Tanaka2013}, making the light curve fainter, redder, and longer-lived \citep{Barnes2013, Grossman2014, Wollaeger2017}.

Neutron star mergers are the likely sources also of short gamma-ray bursts \citep[GRBs][]{Eichler1989, Narayan1992, Fong2013}. Excess emission in the afterglows of some short GRBs has been claimed to be due to kilonovae \citep{Perley2009, Berger2013, Tanvir2013, Yang2015, Jin2016}. 

Here we present a search for electromagnetic counterparts to GW events from the second Advanced Detector Observing Run (O2) using the Las Cumbres Observatory global network of telescopes. We describe the observatory and its unique capabilities in \S2, our follow-up strategy in \S3 and its application to our follow-up of GW170814 \citep{LIGO170814} and GW170817 \citep{LIGO170817,LIGOmma} in \S4. We summarize in \S5. Our followup observations of AT 2017gfo, the optical counterpart of GW170817, are described in companion papers \citep{Arcavi2017, McCully2017}.

\section{Las Cumbres Observatory}

Las Cumbres Observatory (LCO)\footnote{\url{http://lco.global/}} consists of 20 optical telescopes (two 2-meter, nine 1-meter and nine 0.4-meter in diameter) at six sites around the world (Table \ref{tab:lco}), operated robotically using dynamical scheduling software. The observatory capabilities are described in detail in \cite{Brown2013}. Here we summarize the most relevant information.

\begin{deluxetable}{llll}
\tablecaption{The Las Cumbres Observatory global network of robotic telescopes. Each site is identified by a three-letter airport code for brevity.\label{tab:lco}}
\tablehead{\colhead{Observatory} & \colhead{Location} & \colhead{Code} & \colhead{Telescopes}}
\startdata
{McDonald} & {Texas, USA} & {ELP} & {1-m ($\times1$)}\\
{} & {} & {} & {0.4-m ($\times1$)}\\
{Haleakala} & {Hawaii, USA} & {OGG} & {2-m ($\times1$)}\\
{} & {} & {} & {0.4-m ($\times2$)}\\
{El Teide} & {Tenerife, Spain} & {TFN} & {0.4-m ($\times2$)}\\
{CTIO} & {Chile} & {LSC} & {1-m ($\times3$)}\\
{} & {} & {} & {0.4-m ($\times2$)}\\
{Siding Spring} & {Australia} & {COJ} & {2-m ($\times1$)}\\
{} & {} & {} & {1-m ($\times2$)}\\
{} & {} & {} & {0.4-m ($\times2$)}\\
{SAAO} & {South Africa} & {CPT} & {1-m ($\times3$)}\\
\enddata
\end{deluxetable}

Each telescope class uses a different type of imager with a different field of view (FOV) and pixel scale, as listed in Table \ref{tab:imagers}. All imagers are equipped with standard Sloan Digital Sky Survey (SDSS) and Johnson filters, as well as a broad $w$ filter covering the $gri$ bands. The 2-meter telescopes are also equipped with low-resolution (R$\sim400$) Floyds spectrographs.

\begin{deluxetable}{llll}
\tablecaption{Imager properties (including fields of view; FOV) for each class of telescope at LCO.\label{tab:imagers}}
\tablehead{\colhead{Class} & \colhead{Imager} & \colhead{FOV} & \colhead{Pixel Scale (Binning)}}
\startdata
{0.4-m} & {SBIG} & {$29'\times19'$} & {$1.142''$/px ($2\times2$)}\\
{1-m} & {Sinistro} & {$26'\times26'$} & {$0.389''$/px ($1\times1$)}\\
{2-m} & {Spectral} & {$10'\times10'$} & {$0.3''$/px ($2\times2$)}\\
\enddata
\end{deluxetable}

\begin{figure*}[ht!]
\includegraphics[width=\textwidth]{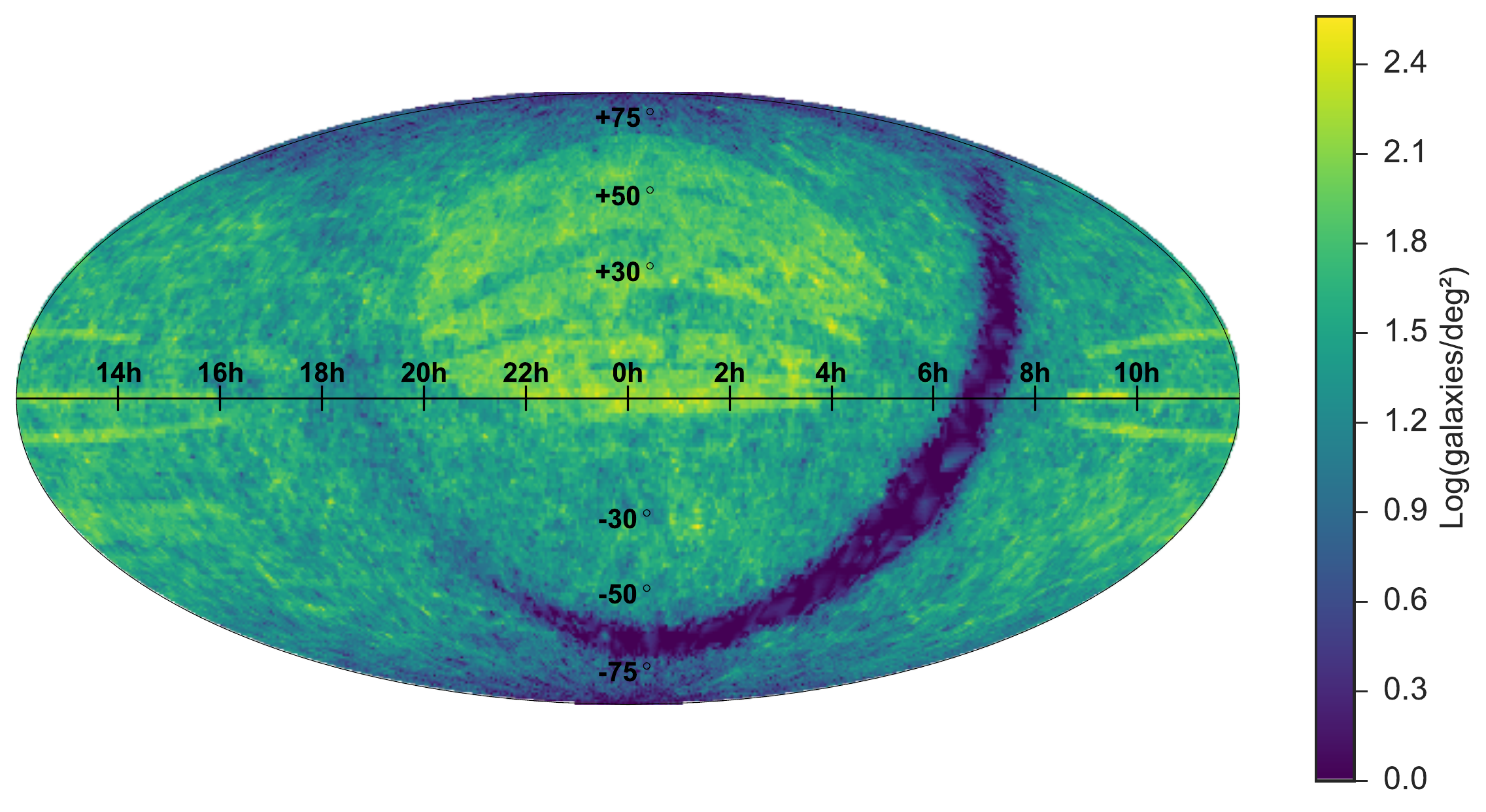}
\caption{\label{fig:glade}Galaxy density (logarithmically in number of galaxies per square degree) in the GLADE catalog shown in a Mollweide projections of R.A. and decl. The sky coverage, set by the surveys that feed into the GLADE catalog, is clearly not uniform (the low density region follows our Galactic plane). However, with over 1.9 million galaxies with distance and $B$-band magnitude estimates, GLADE is the most comprehensive publicly available nearby-galaxy catalog as of O2.}
\end{figure*}

The telescopes are fully robotic and are scheduled by custom software. Users of the observatory submit their requests (which include target information, time constraints, exposure times and the desired telescope class) via web or API\footnote{\url{http://developers.lco.global/}} interfaces. Within minutes, the LCO scheduler automatically assigns the requested observations to a telescope, taking into account the observability of the target, the availability of the different telescopes, and the weather conditions at each site. The schedule is re-evaluated approximately every 15 minutes as existing requests are completed, new requests are submitted, weather shifts, and telescope availability changes. A special ``rapid response'' mode, reserved for the most urgent targets, will stop an ongoing observation to observe a new target as soon as possible (shutter opening is typically within a few minutes from the request being submitted - visibility and weather permitting). The dynamic nature of LCO and its global distribution make it ideal for time-domain astronomy, specifically for quick-response observations of rapidly evolving transients.

\section{The Follow-up Strategy}\label{sec:strategy}

Given the field of view of the LCO imagers, it is not practical to tile an entire GW localization region, which typically ranges in size from tens to hundreds of square degrees (requiring hundreds to thousands of LCO telescope pointings). Instead, we follow the approach presented in \textcite[hereafter \citetalias{G16}]{G16}, which involves targeting only certain galaxies within the GW localization region. 

\citetalias{G16} predicted that the number of galaxies containing $50\%$ of the mass inside a typical O2 GW localization region would be $24\pm6$ (a much more manageable number of pointings compared to tiling the entire localization region). In addition, the Advanced LIGO/Virgo range for mergers involving neutron stars during O2 was $\sim100$\,Mpc \citep{LIGO100Mpc}. At that distance, the peak observed magnitude of the prompt blue emission from kilonova models is $m_{g}\sim21$ \citep{Metzger2015} and the peak of the longer optical/near-infrared (NIR) emission is $m_{i}\sim19-22$ \citep{Barnes2013}. At $>20$\,Mpc, the smallest LCO field of view corresponds to $>60$\,kpc thus encapsulating $>90\%$ of expected merger offsets from their hosts \citep{Fong2013}.

These magnitudes, fields of view, and the relatively small number of pointings expected to cover $50\%$ of the mass motivated us to carry out a GW-LCO follow-up program during LCO semesters 2016B and 2017AB, which overlapped with O2. 

\subsection{The Galaxy Catalog}

\begin{figure*}
\includegraphics[width=\textwidth]{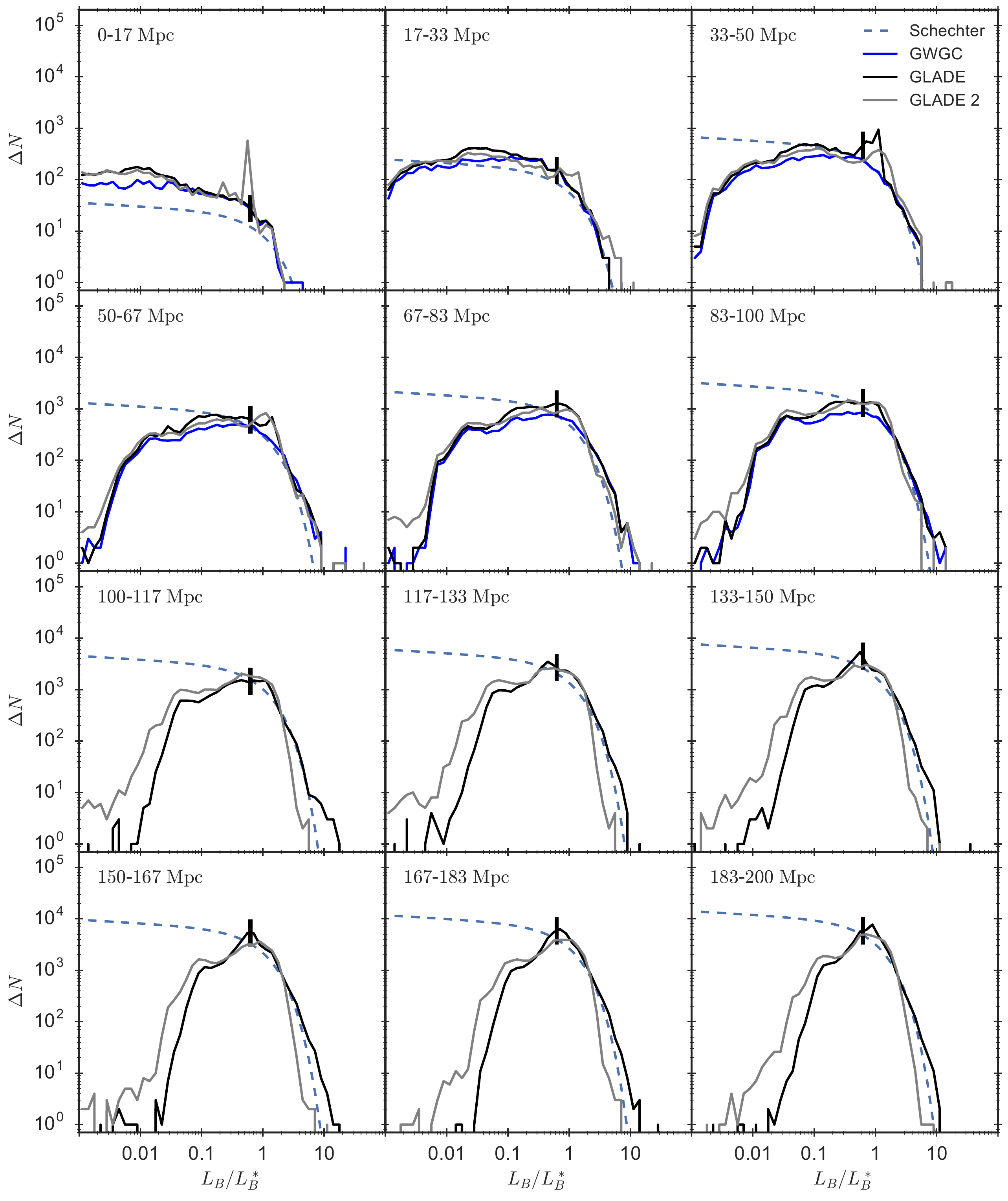}
\caption{\label{fig:schechter}Relative galaxy number density of the GLADE catalogs (version 1 in black, version 2 in gray), the Gravitational Wave Galaxy Catalog (GWGC; blue), and the expected Schechter luminosity function (dashed), corrected for volume, for different distance bins (following \citetalias{G16}). $x_{1/2}$ is marked by a vertical line in each plot. Starting at $\sim30$\,Mpc both GWGC and GLADE miss more and more low-luminosity galaxies; however GLADE is seen to follow the Schechter function quite closely for $L_B/L_B^*>x_{1/2}$ out to $200$\,Mpc (and beyond, see Figure \ref{fig:completeness}).}
\end{figure*}

We use Version 1 of the Galaxy List for the Advanced Detector Era (GLADE; Fig. \ref{fig:glade}) catalog complied by \cite{Dalya2016}\footnote{\url{http://aquarius.elte.hu/glade/}}. It contains 1,918,147 galaxies amassed from the Gravitational Wave Galaxy Catalog \citep[GWGC][]{White2011}, the 2MASS XSC \citep{Skrutskie2006}, the 2MPZ \citep{Bilicki2014}, and the HyperLEDA \citep{Makarov2014} catalogs (Fig. \ref{fig:glade}). An apparent $B$-band magnitude is associated with every galaxy from either direct measurement or by deduction from other available magnitudes. GLADE also contains distance information for each galaxy (compiled from various sources). Version 2 of the GLADE catalog, containing more than 3.6 million galaxies, was made available during O2. However, compared to Version 1, most of the added galaxies do not have distance or $B$-band magnitude estimates. We thus chose to continue using Version 1 for the entire O2 run, though we analyze the completeness of both versions below. GLADE is the largest census of the nearby Universe that was publicly available during O2.

\begin{figure}
\includegraphics[width=\columnwidth]{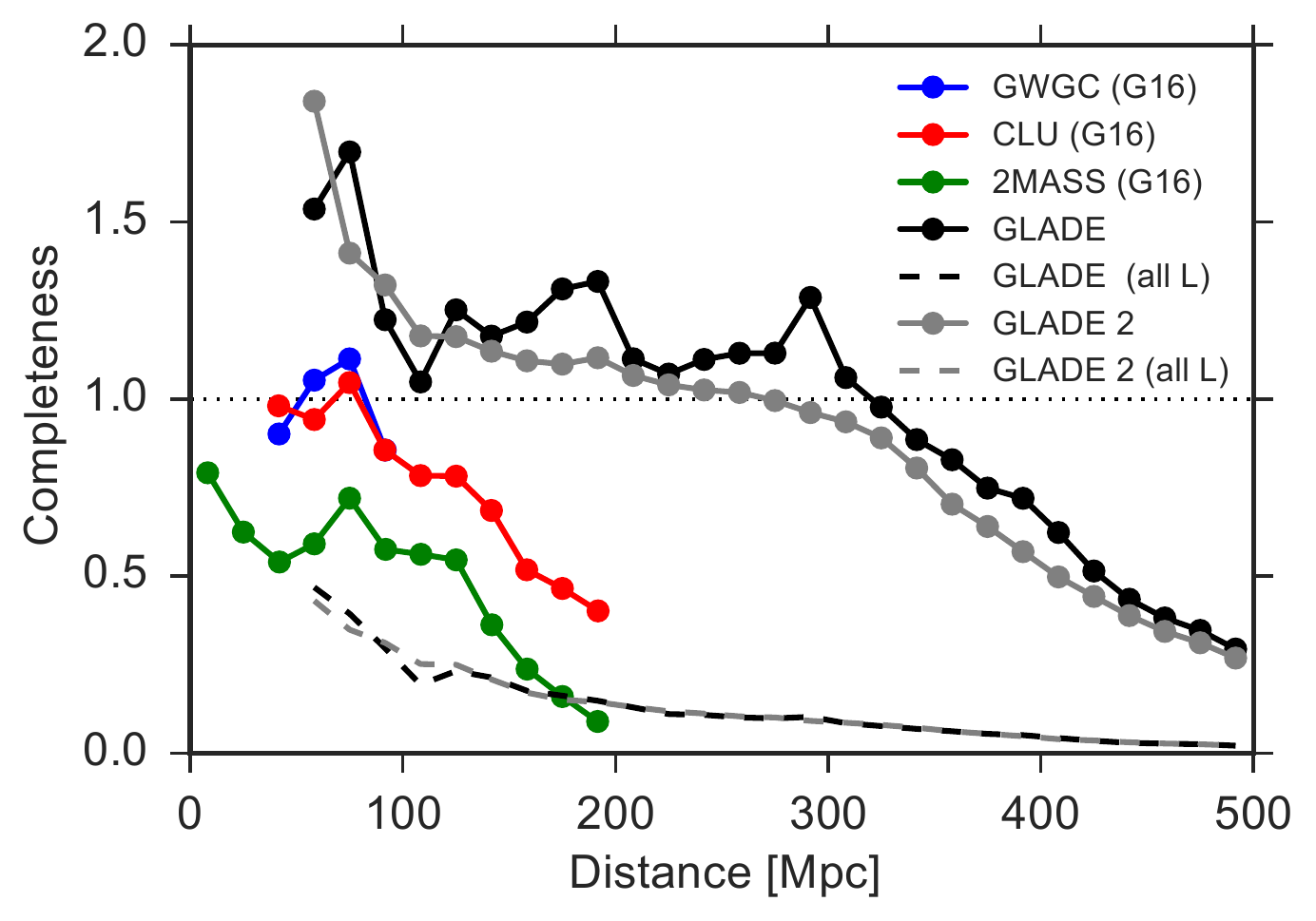}
\caption{\label{fig:completeness}Completeness of GLADE (calculated from Figure \ref{fig:schechter}) relative to the Schechter function for galaxies with $L_B/L_B^*>x_{1/2}$ (i.e. galaxies brighter than the median galaxy luminosity; solid black and gray lines), and for all galaxies (dashed black and gray lines). We also present the data for other catalogs from \citetalias{G16}. The greater than $100\%$ completeness for $L_B/L_B^*>x_{1/2}$ galaxies in GLADE at low distances is mostly due to the overabundances of galaxies seen in Figure \ref{fig:schechter} at these distances. GLADE is complete for $L_B/L_B^*>x_{1/2}$ out to $\sim300$\,Mpc.}
\end{figure}

Following \citetalias{G16}, we analyze the completeness of the GLADE catalog relative to the Schechter luminosity function \citep{Schechter1976}, which provides a form for the number density of galaxies $\rho_{\rm gal}\left(x\right)dx=\phi^{*}x^{\alpha}e^{-x}dx$, where $x=L/L^*$ with $L$ the luminosity of the galaxy and $L^*$ a parameter of the function. Since the GLADE catalog contains $B$-band data, we use $x=L_B/L_B^*$. In order to compare the GLADE catalog with those presented in \citetalias{G16}, we adopt the same parameters as they did, namely: $\phi^*=1.6\times10^{-2}h^3$\,Mpc$^{-3}$ (with $h=0.7$), $\alpha=-1.07$ and an $L_B^*$ corresponding to $M_B^*=-20.47$. We reproduce Figure 2 from G16, for the GLADE catalog and the GWGC for comparison, in Fig. \ref{fig:schechter} here. This figure shows the relative number density of galaxies in GLADE vs. the Schechter function for various distance bins. We also adopt $x_{1/2}=0.626$ as the median of the luminosity function (i.e. half of the total luminosity is in galaxies above this value and half is in galaxies below it).

While both GWGC and GLADE are missing low-luminosity galaxies at distances greater than $\sim30$\,Mpc, both are complete relative to the Schechter function for galaxies more luminous than $x_{1/2}$ (GWGC contains galaxies only out to $100$\,Mpc, while GLADE extends farther). This is the original motivation for focusing on the top $50\%$ of the luminosity (or mass) distributions: galaxy catalogs are complete for galaxies brighter than $x_{1/2}$ out to the relevant distances for GW NS--NS and NS--BH detections. In fact, we find that GLADE is complete out to $\sim300$\,Mpc for such galaxies (Fig. \ref{fig:completeness}). In addition, NS--NS mergers are expected in galaxies with a $B$-band luminosity of roughly $L_B^*$, since most short GRBs have been seen to occur in such galaxies \citep{Berger2014}.

We find some peculiar overabundances of galaxies in both versions of the GLADE catalog for $L_B/L_B^*\approx1$ at $33$--$67$\,Mpc and in version 2 of the GLADE catalog near $x_{1/2}$ at $<17$\,Mpc. These overabundances lead to $>100\%$ completeness values compared to the Schechter function at those distances in Figure \ref{fig:completeness}, and may be due to artifacts in the GLADE catalog.

\subsection{Galaxy Prioritization}\label{sec:algorithm}

For maximizing the efficiency of optical follow-up observations of GW triggers, we wish to prioritize galaxies that are in higher probability regions of the GW localization and which are more massive (assuming compact object mergers follow the mass distribution). Everything else being equal among those, we will prefer galaxies that are closer to us, in which a possible counterpart is more likely to be detected. This is a slightly different approach than the one outlined by \cite{Singer2016}.

We include only galaxies that are inside the $99\%$ GW localization region and less than $3\sigma$ away from the GW distance estimate (these criteria are relaxed to $99.995\%$ and $5\sigma$ if the original cut leaves no galaxies). Second, we remove galaxies that are fainter than $x_{1/2}$ (or a lower threshold to make sure that at least $100$ galaxies remain) based on a Schechter function with $\alpha = -1.07$ and an $L^{*}$ corresponding to $M_B=-20.7$ (this is similar but not identical to the magnitude of -20.47 used by \citetalias{G16}).

Of the galaxies that remain after the position, distance, and luminosity cuts, each is given three scores (which we detail below) based on
\begin{enumerate}
\item its location in the GW localization region (including distance information), $S_{\rm loc}$,
\item its absolute $B$-band luminosity (as an indicator of mass), $S_{\rm lum}$, and
\item the likelihood of detecting a counterpart at its distance, $S_{\rm det}$.
\end{enumerate}

The localization information provided by the GW detectors includes a probability for each position of the sky, so that the probability of the true GW source to be at a certain R.A. and decl. is a given  $p_{\rm pos}\left({\rm R.A.,decl.}\right)$. The localization also includes a mean distance estimate $\mu_{\rm dist}$, standard deviation $\sigma_{\rm dist}$, and normalization $N_{\rm dist}$ per R.A. and decl. We assume that the distance estimate probability function is a Gaussian with the provided mean, standard deviation, and normalization, so that the probability of the source being at distance $D$ for a certain R.A. and decl. is:
\begin{equation}
p_{\rm dist}\left({\rm R.A.,decl.,}D\right) = N_{\rm dist}\left({\rm R.A.,decl.}\right){\cdot}e^{-\frac{\left[D-\mu_{\rm dist}\left({\rm R.A.,decl.}\right)\right]^2}{2\sigma_{\rm dist}^2\left({\rm R.A.,decl.}\right)}}.
\end{equation}
The location score of a galaxy at a certain RA, decl. and distance $D$ is then
\begin{equation}
S_{\rm loc} = p_{\rm pos}\left({\rm R.A.,decl.}\right){\times}p_{\rm dist}\left({\rm R.A.,decl.,}D\right).
\end{equation}

We then calculate the $B$-band luminosity of the galaxy, $L_B$ (based on the $B$-band magnitude and distance provided in the GLADE catalog), and assign it a score
\begin{equation}
S_{\rm lum} = \frac{L_B}{\sum{L_B}},
\end{equation}
where the sum is over all of the galaxies being considered.

Finally, we score each galaxy on the likelihood of detecting a counterpart there. We assume a limiting magnitude for our exposures, $m_{\rm lim}$, and convert it to a limiting luminosity at the distance of each galaxy, $L_{\rm lim}$. We also define the likely counterpart magnitude range, $M_{\rm KN,min}$--$M_{\rm KN,max}$ and convert those magnitudes to luminosities. We then give each galaxy a detection likelihood score,
\begin{equation}
S_{\rm det} = \frac{L_{\rm KN,max}-L_{\rm lim}}{L_{\rm KN,max}-L_{\rm KN,min}}
\end{equation}
while limiting it to being between $0.01$ and $1$. So, for example, a galaxy for which our limiting luminosity is equal to or fainter than the minimum luminosity we expect from the counterpart (i.e. we are guaranteed to see it) receives a detectability score of $1$, while a galaxy for which our limiting magnitude is equal to or brighter than the maximum luminosity we expect from the counterpart (i.e. we will not see it) receives a detectability score of $0.01$ (we avoid giving it a score of $0$ in order to not exclude distant galaxies completely). We use a conservative selection of parameters ($m_{\rm lim} = 22$, $M_{\rm KN,min} = -17$ and $M_{\rm KN,max} = -12$), making this score quite high for most galaxies in the NS--NS and NS--BH O2 detectability range of LIGO/Virgo. This criterion therefore has no effect on very close events, and will only slightly prefer closer galaxies in events around $\sim100$\,Mpc. 

The product of these three scores is the final score assigned to each galaxy,
\begin{equation}
S = S_{\rm loc}{\times}S_{\rm lum}{\times}S_{\rm det}.
\end{equation}
This score is then used to prioritize which galaxies to observe following a trigger. In Section \ref{sec:G298048} we show that this prioritization procedure ranked the correct host galaxy of a GW source as fifth out of the entire GLADE catalog.

\subsection{The Triggering Process}

\begin{figure}
\includegraphics[width=\columnwidth]{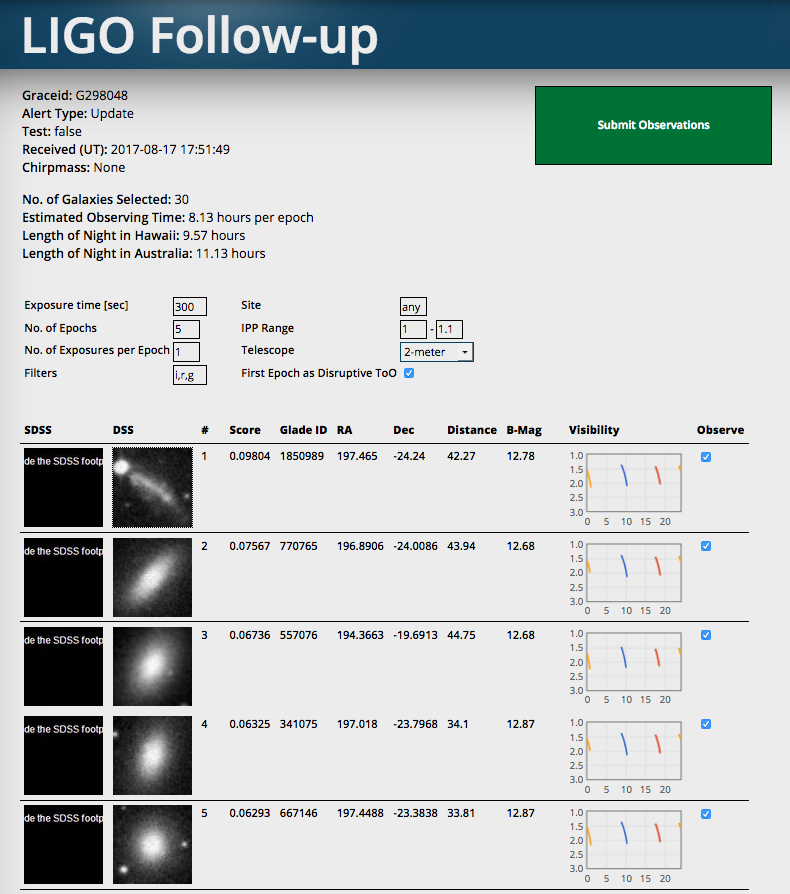}
\caption{\label{fig:gui}Screenshot of the web interface used for verifying the galaxies selected for monitoring following a GW trigger, ordered in descending priority (only the top 5 galaxies of 100 displayed are shown here). The galaxies to be observed and the observing parameters can be modified by the user before submitting the observations to the LCO scheduler. This screenshot shows the webpage generated following the G298048 trigger for GW170817. The galaxy containing the optical counterpart can be seen in the list (galaxy number 5). In principle, these fields can be sent to the LCO telescopes automatically within seconds of the alert being received, without any human intervention.}
\end{figure}

\begin{figure*}[ht!]
\includegraphics[width=\textwidth]{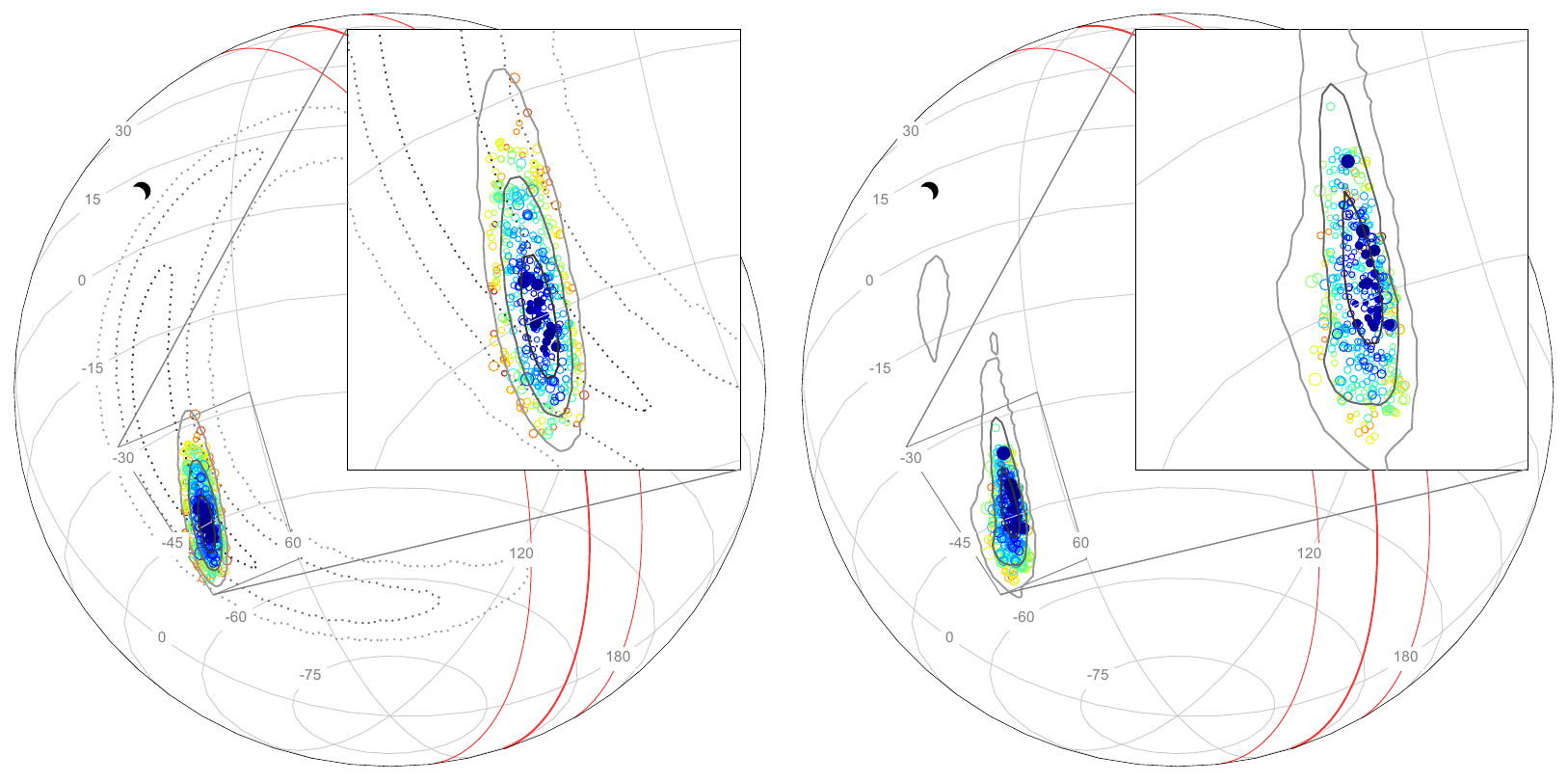}
\caption{\label{fig:G297595}Localization region (contours) and the matched galaxies (circles) for G279595 (observed galaxies are in filled circles). The orthographic projection on the left is for the initial localization (LIGO only shown in dashed lines, LIGO/Virgo in solid lines), while the one on the right is for the updated localization. The contours indicate $50\%$, $90\%$, and $99\%$ confidence bounds. The colors of the circles denote the priority of the galaxies (low priority in yellow, high priority in blue). The position of the plane of the Milky Way is indicated in red lines, with a $\pm10^{\circ}$-wide band. The position of the moon at the time of the trigger is indicated with a crescent symbol. Insets show a more detailed view of each localization.}
\end{figure*}

We employ a GCN listener, based on \texttt{pyGCN}\footnote{\url{https://github.com/lpsinger/pygcn}}, to receive LIGO/Virgo alerts via VOEvent \citep{Seaman2011}, ingest them to a database, download the \cite{SingerPrice2016} HEALPIX localization map (which includes a distance constraint) attached to the alert, cross-check that localization with the GLADE galaxy catalog, and prioritize the galaxies to be observed according to the algorithm described above\footnote{\url{https://github.com/svasyly/pygcn}}. This process takes a few seconds, after which the top galaxies on the list can be sent to the LCO scheduler automatically. The observing requests use Intra-Proposal Priority (IPP) values proportional to the priorities determined for the galaxies. IPP is used by the LCO scheduler to resolve scheduling conflicts if not all targets can be observed\footnote{\url{https://lco.global/files/User_Documentation/the_new_priority_factor.pdf}}.

During O2 we take the precaution of having humans verify the candidate galaxies to be observed before the triggers are delivered to the LCO scheduler. This verification step is done through a webpage (Fig. \ref{fig:gui}) which displays the top 100 galaxies selected sorted by priority, with the first $30$ galaxies automatically selected. For each galaxy we present an SDSS cutout image, if available, a Digital Sky Survey (DSS) image, and the observability of the galaxy from the various LCO sites. The user can change the selection of galaxies and compare the total estimated observing time needed for each selection to the available length of night time at two representative LCO sites (Australia and Hawaii). The default exposure sequence is 300 seconds in each of the $g$,$r$ and $i$ filters to be taken with the 2-meter telescopes. This exposure time was chosen in order to reach a signal to noise of $10$ at the expected magnitudes listed above for the different emission mechanisms for a kilonova at $100$\,Mpc. The number of galaxies selected by default ($30$) is the typical number that could fit in a full observing night given these exposure times, and is also the amount predicted by \citetalias{G16} to contain roughly $50\%$ of the mass in the localization region. However, the user can change the exposure times, numbers, filters, and telescope class based on the specific trigger parameters. The user can also select whether to submit the first epoch as a rapid response observation (which interrupts observations that were ongoing at the time of the trigger). 

Once the galaxies, exposure sequences, number of epochs, and telescope class are selected, the information is converted into observing sequences which are then submitted programmatically to the LCO scheduler using its API. After the images are taken, they are automatically ingested and processed by the \texttt{LCOGTSNpipe} pipeline \citep{Valenti2016} and displayed on a webpage for manual scanning, next to SDSS (if available) and DSS images of the field for comparison. Image subtraction can then be performed in order to detect faint transients using SDSS templates when available or subtracting the different LCO epochs off of each other to search for changing sources, otherwise.

\section{Advanced LIGO/Virgo Observing Run 2}

O2 ran from 2016 November 30 to 2017 August 25, with Virgo joining the two LIGO detectors on 2017 August 01 (UT used throughout). Both LIGO detectors were taken offline on 2017 May 08 for commissioning activities, with the Livingston detector resuming operations on 2017 May 26 and the Hanford one on 2017 June 08. Several triggers were issued for follow-up to the EM community. Here we detail our follow-up observations for two such triggers.

\subsection{G297595 / GW170814}\label{sec:G297595}

G297595 was the first event detected by both LIGO detectors and the Virgo detector in real time, with the Virgo detection contributing significantly to the localization \citep{LIGO170814}. The Virgo detection decreased the $50\%$ ($90\%$) localization region from 333\,deg$^2$ (1158\,deg$^2$) to 22\,deg$^2$ (97\,deg$^2$). The signal was identified on 2017 August 14 10:30:43 with a very low false-alarm rate \citep[$\sim1$ per $80,000$ years][]{LVC170814initGCN}, as a likely BH-BH merger at $\sim500$\,Mpc. Despite the lack of a NS component and the large distance, we triggered our follow-up program given the relatively small localization region. 

Following the VOEvent trigger sent at 2017 August 14 11:01:49, and the circular issued at 12:28:42 \citep{LVC170814initGCN}, we triggered 2-meter follow-up imaging of 30 galaxies at 15:18:43 (Fig. \ref{fig:G297595}, left panel). On 2017 August 16 07:02:19 an updated localization was issued. The region moved east and grew slightly to 36\,deg$^2$ (190\,deg$^2$) for the $50\%$ ($90\%$) localization probability, due to marginalization over calibration uncertainties \citep{LVC170814updateGCN}. The updated localization was sent by VOEvent at 17:01:54. We stopped all of our ongoing observation requests at 21:37:35 and re-submitted a new set of 30 galaxies based on the updated localization at 21:39:24 (Fig. \ref{fig:G297595}, right panel). In total, 63 images were obtained for 20 galaxies, 16 from the original localization (Table \ref{tab:G297595_initial}; the initial measurements of which were reported in \citealt{HosseinzadehGCN}) and four from the updated localization (Table \ref{tab:G297595_updated}). Upper limits are calculated by calibrating a local sequence of stars in each field to the AAVSO Photometric All-Sky Survey (APASS) catalog \citep{Henden2009} and are presented in the AB system with no extinction corrections applied. 

No obvious optical counterparts were detected. All observations from this trigger were stopped on 2017 August 17 23:04:02 in order to free the telescopes to aggressively pursue the next trigger.

\begin{deluxetable*}{rDDcDDlllD}
\tablecaption{LCO follow-up observations of the inital LIGO/Virgo localization for trigger G297595 in descending order of galaxy priority. The leftmost four columns are provided as-is from the GLADE catalog. A limiting magnitude was not calculated for fields with very few APASS stars visible. All exposures were 300 seconds long. See Table \ref{tab:lco} for the list of site abbreviations used in the telescope column.\label{tab:G297595_initial}}
\tablehead{\colhead{GLADE} & \multicolumn2c{RA} & \multicolumn2c{Dec} & \colhead{Distance} & \multicolumn2c{m$_{\rm B}$} & \multicolumn2c{$L_{\rm B}/L_{\rm B}^*$} & \colhead{UT} & \colhead{Telescope} & \colhead{Filt.} & \multicolumn2c{Limiting} \\
\colhead{ID} & \multicolumn2c{} & \multicolumn2c{} & \colhead{[Mpc]} & \multicolumn2c{} & \multicolumn2c{} & & & & \multicolumn2c{Mag. ($3\sigma$)}}
\decimals
\startdata
723415 & 40.523499 & -45.337452 & 432 & 15.57 & 5.808 & 2017-08-14 16:46:38 & COJ 2m & g & 21.52 \\
723415 & 40.523499 & -45.337452 & 432 & 15.57 & 5.808 & 2017-08-14 16:52:10 & COJ 2m & r & 21.17 \\
723415 & 40.523499 & -45.337452 & 432 & 15.57 & 5.808 & 2017-08-14 16:57:41 & COJ 2m & i & 20.41 \\
721389 & 42.015915 & -44.111561 & 476 & 16.09 & 4.365 & 2017-08-14 16:00:52 & COJ 2m & g & 21.35 \\
721389 & 42.015915 & -44.111561 & 476 & 16.09 & 4.365 & 2017-08-14 16:08:39 & COJ 2m & r & 21.41 \\
721389 & 42.015915 & -44.111561 & 476 & 16.09 & 4.365 & 2017-08-14 16:14:12 & COJ 2m & i & 21.66 \\
787654 & 40.73167 & -44.360573 & 442 & 16.18 & 3.467 & 2017-08-14 17:06:51 & COJ 2m & g & 20.77 \\
787654 & 40.73167 & -44.360573 & 442 & 16.18 & 3.467 & 2017-08-14 17:12:22 & COJ 2m & r & 18.89 \\
787654 & 40.73167 & -44.360573 & 442 & 16.18 & 3.467 & 2017-08-14 17:17:53 & COJ 2m & i & 21.56 \\
556821 & 41.190659 & -45.095711 & 440 & 16.63 & 2.270 & 2017-08-14 18:20:21 & COJ 2m & g & 20.26 \\
556821 & 41.190659 & -45.095711 & 440 & 16.63 & 2.270 & 2017-08-14 18:26:34 & COJ 2m & r & 18.99 \\
556821 & 41.190659 & -45.095711 & 440 & 16.63 & 2.270 & 2017-08-14 18:32:06 & COJ 2m & i & 17.48 \\
625999 & 41.654194 & -42.367088 & 307 & 14.42 & 8.472 & 2017-08-15 14:28:10 & OGG 2m & g & 21.82 \\
625999 & 41.654194 & -42.367088 & 307 & 14.42 & 8.472 & 2017-08-15 14:34:20 & OGG 2m & r & 22.01 \\
625999 & 41.654194 & -42.367088 & 307 & 14.42 & 8.472 & 2017-08-15 14:39:52 & OGG 2m & i & 21.82 \\
625999 & 41.654194 & -42.367088 & 307 & 14.42 & 8.472 & 2017-08-15 15:08:26 & OGG 2m & g & 21.36 \\
706152 & 40.122471 & -45.386868 & 494 & 16.78 & 2.489 & 2017-08-15 13:30:24 & COJ 2m & g & 20.38 \\
706152 & 40.122471 & -45.386868 & 494 & 16.78 & 2.489 & 2017-08-15 13:35:57 & COJ 2m & r & 20.53 \\
706152 & 40.122471 & -45.386868 & 494 & 16.78 & 2.489 & 2017-08-15 13:41:28 & COJ 2m & i & 20.22 \\
752527 & 42.434814 & -42.327412 & 322 & 15.05 & 5.200 & 2017-08-15 14:47:23 & OGG 2m & g & 0.00 \\
752527 & 42.434814 & -42.327412 & 322 & 15.05 & 5.200 & 2017-08-15 14:52:54 & OGG 2m & r & 0.00 \\
752527 & 42.434814 & -42.327412 & 322 & 15.05 & 5.200 & 2017-08-15 14:58:25 & OGG 2m & i & 0.00 \\
1066576 & 41.292435 & -46.59705 & 398 & 16.49 & 2.113 & 2017-08-14 17:27:53 & COJ 2m & g & 21.37 \\
1066576 & 41.292435 & -46.59705 & 398 & 16.49 & 2.113 & 2017-08-14 17:33:24 & COJ 2m & r & 21.50 \\
1066576 & 41.292435 & -46.59705 & 398 & 16.49 & 2.113 & 2017-08-14 17:38:56 & COJ 2m & i & 21.52 \\
1005823 & 42.07933 & -44.054893 & 450 & 16.89 & 1.863 & 2017-08-15 14:10:58 & COJ 2m & g & 20.53 \\
1005823 & 42.07933 & -44.054893 & 450 & 16.89 & 1.863 & 2017-08-15 14:16:30 & COJ 2m & r & 20.65 \\
1005823 & 42.07933 & -44.054893 & 450 & 16.89 & 1.863 & 2017-08-15 14:22:02 & COJ 2m & i & 20.34 \\
1031304 & 41.217548 & -46.500435 & 401 & 16.63 & 1.893 & 2017-08-15 14:53:00 & COJ 2m & g & 20.87 \\
1031304 & 41.217548 & -46.500435 & 401 & 16.63 & 1.893 & 2017-08-15 14:58:31 & COJ 2m & r & 20.84 \\
1031304 & 41.217548 & -46.500435 & 401 & 16.63 & 1.893 & 2017-08-15 15:04:03 & COJ 2m & i & 20.63 \\
622864 & 41.392635 & -44.539589 & 316 & 15.47 & 3.404 & 2017-08-14 19:12:39 & COJ 2m & g & 20.45 \\
622864 & 41.392635 & -44.539589 & 316 & 15.47 & 3.404 & 2017-08-14 19:18:10 & COJ 2m & r & 20.17 \\
622864 & 41.392635 & -44.539589 & 316 & 15.47 & 3.404 & 2017-08-14 19:23:41 & COJ 2m & i & 19.91 \\
1415752 & 42.418404 & -44.226032 & 421 & 16.77 & 1.832 & 2017-08-15 13:49:58 & COJ 2m & g & 20.43 \\
1415752 & 42.418404 & -44.226032 & 421 & 16.77 & 1.832 & 2017-08-15 13:55:31 & COJ 2m & r & 20.59 \\
1415752 & 42.418404 & -44.226032 & 421 & 16.77 & 1.832 & 2017-08-15 14:01:02 & COJ 2m & i & 20.27 \\
1181112 & 41.086727 & -43.939903 & 452 & 16.86 & 1.930 & 2017-08-15 14:32:00 & COJ 2m & g & 20.76 \\
1181112 & 41.086727 & -43.939903 & 452 & 16.86 & 1.930 & 2017-08-15 14:37:31 & COJ 2m & r & 20.75 \\
1181112 & 41.086727 & -43.939903 & 452 & 16.86 & 1.930 & 2017-08-15 14:43:03 & COJ 2m & i & 20.59 \\
806902 & 42.062664 & -45.159618 & 446 & 16.93 & 1.770 & 2017-08-14 16:25:24 & COJ 2m & g & 20.79 \\
806902 & 42.062664 & -45.159618 & 446 & 16.93 & 1.770 & 2017-08-14 16:30:55 & COJ 2m & r & 18.41 \\
806902 & 42.062664 & -45.159618 & 446 & 16.93 & 1.770 & 2017-08-14 16:36:27 & COJ 2m & i & 20.71 \\
1647694 & 42.0008 & -46.31986 & 381 & 16.24 & 2.443 & 2017-08-14 18:52:50 & COJ 2m & g & 21.47 \\
1647694 & 42.0008 & -46.31986 & 381 & 16.24 & 2.443 & 2017-08-14 18:59:02 & COJ 2m & r & 21.52 \\
1647694 & 42.0008 & -46.31986 & 381 & 16.24 & 2.443 & 2017-08-14 19:04:33 & COJ 2m & i & 20.50 \\
62667 & 41.574863 & -44.984894 & 487 & 17.22 & 1.619 & 2017-08-15 15:14:03 & COJ 2m & g & 20.78 \\
62667 & 41.574863 & -44.984894 & 487 & 17.22 & 1.619 & 2017-08-15 15:19:34 & COJ 2m & r & 20.88 \\
62667 & 41.574863 & -44.984894 & 487 & 17.22 & 1.619 & 2017-08-15 15:25:06 & COJ 2m & i & 20.54 \\
\enddata
\end{deluxetable*}

\begin{deluxetable*}{rDDcDDlllD}
\tablecaption{LCO follow-up observations of the updated LIGO/Virgo localization for trigger G297595 in descending order of galaxy priority. The leftmost four columns are provided as-is from the GLADE catalog. All exposures were 300 seconds long. See Table \ref{tab:lco} for the list of site abbreviations used in the telescope column.\label{tab:G297595_updated}}
\tablehead{\colhead{GLADE} & \multicolumn2c{RA} & \multicolumn2c{Dec} & \colhead{Distance} & \multicolumn2c{m$_{\rm B}$} & \multicolumn2c{$L_{\rm B}/L_{\rm B}^*$} & \colhead{UT} & \colhead{Telescope} & \colhead{Filt.} & \multicolumn2c{Limiting Mag.} \\
\colhead{ID} & \multicolumn2c{} & \multicolumn2c{} & \colhead{[Mpc]} & \multicolumn2c{} & \multicolumn2c{} & & & & \multicolumn2c{($3\sigma$)}}
\decimals
\startdata
789138 & 49.259586 & -40.445934 & 414 & 14.86 & 10.280 & 2017-08-17 17:37:02 & COJ 2m & g & 21.75 \\
789138 & 49.259586 & -40.445934 & 414 & 14.86 & 10.280 & 2017-08-17 17:44:49 & COJ 2m & r & 21.86 \\
789138 & 49.259586 & -40.445934 & 414 & 14.86 & 10.280 & 2017-08-17 17:50:20 & COJ 2m & i & 21.60 \\
632134 & 49.61784 & -42.020428 & 420 & 15.30 & 7.047 & 2017-08-17 16:18:03 & COJ 2m & g & 21.95 \\
632134 & 49.61784 & -42.020428 & 420 & 15.30 & 7.047 & 2017-08-17 16:23:33 & COJ 2m & r & 21.94 \\
632134 & 49.61784 & -42.020428 & 420 & 15.30 & 7.047 & 2017-08-17 16:29:04 & COJ 2m & i & 21.59 \\
1385568 & 45.634289 & -46.345387 & 455 & 16.66 & 2.353 & 2017-08-17 13:41:49 & COJ 2m & g & 21.22 \\
1385568 & 45.634289 & -46.345387 & 455 & 16.66 & 2.353 & 2017-08-17 13:47:22 & COJ 2m & r & 21.47 \\
1385568 & 45.634289 & -46.345387 & 455 & 16.66 & 2.353 & 2017-08-17 17:57:11 & COJ 2m & g & 21.87 \\
1385568 & 45.634289 & -46.345387 & 455 & 16.66 & 2.353 & 2017-08-17 18:02:42 & COJ 2m & r & 22.19 \\
1385568 & 45.634289 & -46.345387 & 455 & 16.66 & 2.353 & 2017-08-17 18:08:14 & COJ 2m & i & 21.77 \\
712985 & 48.530678 & -42.086346 & 440 & 16.17 & 3.467 & 2017-08-17 14:43:49 & OGG 2m & g & 21.62 \\
712985 & 48.530678 & -42.086346 & 440 & 16.17 & 3.467 & 2017-08-17 14:49:22 & OGG 2m & r & 22.41 \\
712985 & 48.530678 & -42.086346 & 440 & 16.17 & 3.467 & 2017-08-17 14:54:52 & OGG 2m & i & 21.98 \\
\enddata
\end{deluxetable*}

\subsection{G298048 / GW170817}\label{sec:G298048}

\begin{figure*}
\includegraphics[width=\textwidth]{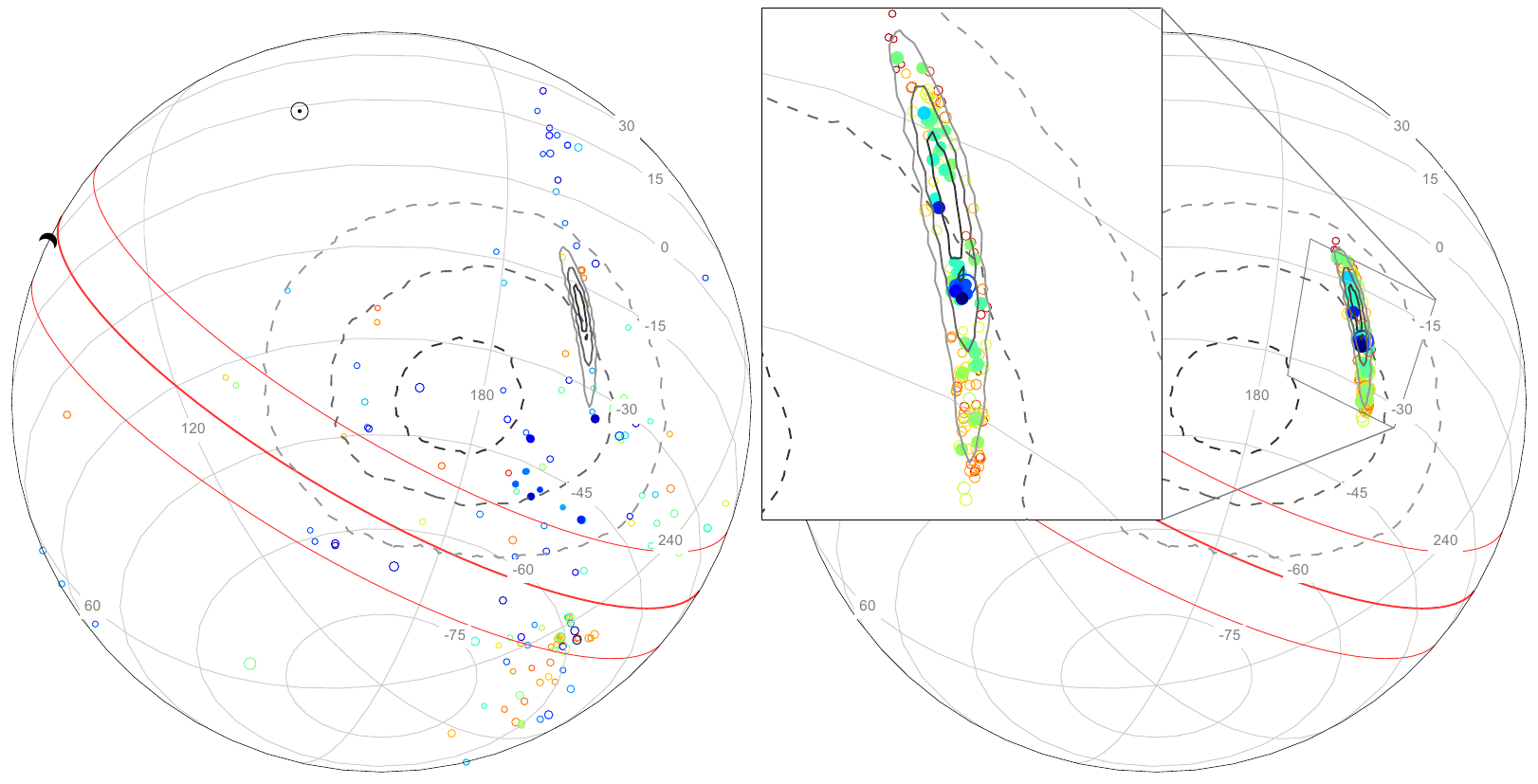}
\caption{\label{fig:G298048}Localization region (contours) and the matched galaxies (circles) for G298048. The orthographic projection on the left shows the galaxies selected for observations (filled circles) following the initial LIGO localization (which covered most of the sky; not shown) and Fermi localization (dashed contours), while the one on the right is for the updated LIGO/Virgo localization (solid contours). The contours indicate $50\%$, $90\%$, and $99\%$ confidence bounds. The colors of the circles denote the priority of the galaxies (low priority in yellow, high priority in blue). The position of the plane of the Milky Way is indicated in red lines, with a $\pm10^{\circ}$-wide band. The position of the moon is indicated with the a crescent symbol, and that of the sun with a $\odot$ symbol. The inset shows a more detailed view of the LIGO/Virgo localization. The galaxy containing the optical counterpart is marked with an additional circle around it.}.
\end{figure*}

\begin{figure}
\includegraphics[width=\columnwidth]{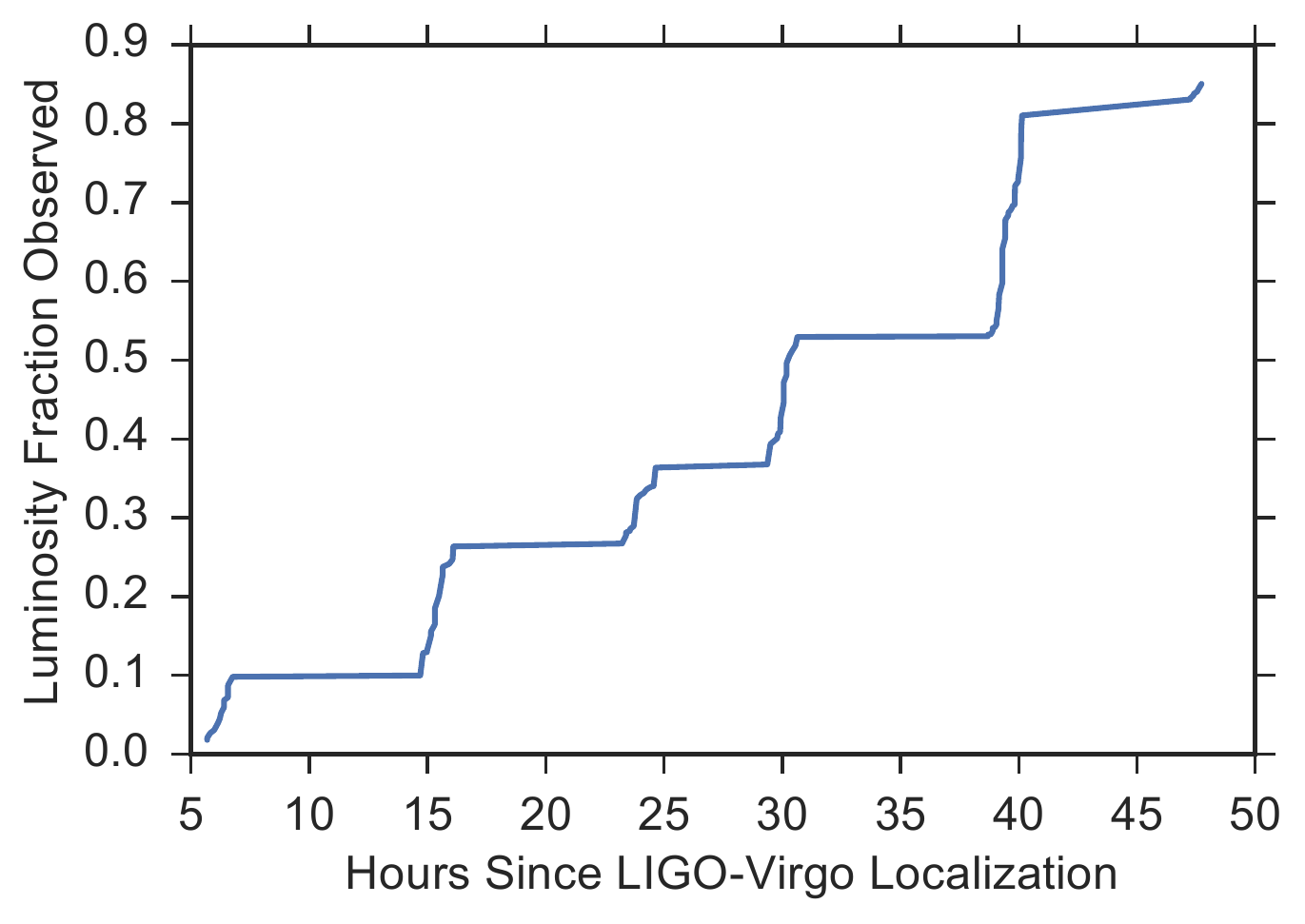}
\caption{\label{fig:obs_f}Cumulative luminosity observed as a fraction of the total luminosity contained in the 182 galaxies identified by our algorithm in the LIGO/Virgo localization region of GW170817. Due to the short visibility window of the localization region (only $\sim$2 hours per night), it took approximately 30 hours to cover half of the luminosity. Given the importance of the trigger as the first NS-NS detection, we continued to observe additional galaxies until we reached $85\%$ of the total galaxy luminosity.}
\end{figure}

This event \citep{LIGO170817, LIGOmma} was detected as a weak $\gamma$-ray transient, interpreted as a likely short GRB, by the Fermi Gamam-ray Burst Monitor (GBM) on 2017 August 17 12:41:06 \citep{FermiGCN} and then associated with a low false-alarm rate ($\sim1$ in $10,000$ years), likely NS--NS merger GW event detected two seconds earlier in the LIGO Hanford detector \citep{LVC170817initGCN}. The GBM localization was distributed at 13:47:37 via GCN Circular before a full LIGO/Virgo localization was available. The initial GW localization (solely from the Hanford detector, thus covering most of the sky, with a preliminary distance estimate) was distributed via VOEvent at 13:08:16. We triggered a first set of $11$ galaxies selected from the GW localization, chosen to lie near the GBM localization center for LCO 1-m imaging starting at 16:34:34 (Fig. \ref{fig:G298048}, left panel). We obtained a total of $30$ images of eight galaxies (Table \ref{tab:Fermi_followup}), which we immediately inspected for transients and found none.

A LIGO/Virgo localization, consistent with the GBM region but offset from its center, was distributed by VOEvent at 17:51:48. A total of 182 galaxies were identified by our algorithm in the LIGO/Virgo localization region (adhering to the criteria detailed in Section \ref{sec:algorithm}). We triggered LCO 0.4-m and 1-m imaging of the top 60 galaxies in the prioritized galaxy list starting at 22:36:19 (Fig. \ref{fig:G298048}, right panel), and stopped observations of the previous GBM-selected galaxies shortly thereafter. Given the limited observability of the updated GW localization region to the first $1.5$--$2$ hours of the night, we decided to trigger single 300-second $w$-band exposures in order to be able to cover many galaxies in this short window. We obtained a total of $96$ images of the triggered $60$ galaxies (Table \ref{tab:LVC_followup}), which constitute $85\%$ of the luminosity contained in the 182 identified galaxies (Fig. \ref{fig:obs_f}). 

The optical counterpart, AT 2017gfo, announced by \cite{CoulterGCN} was in one of our observed galaxies, ranked fifth in priority. The transient is clearly present in our imaging of that galaxy, beginning with our first epoch \citep{ArcaviGCN} taken approximately one hour before the \cite{CoulterGCN} announcement. We continued to monitor other galaxies in case AT 2017gfo turned out to be an unrelated transient, but as it became clear that the color and luminosity of AT 2017gfo were evolving rapidly, we gradually shifted towards observing just that position with more bands and more telescopes. No obvious counterpart candidates were found in the other galaxies surveyed. The follow-up data obtained for AT 2017gfo is the topic of two companion papers \citep{Arcavi2017, McCully2017}.

\section{Summary}

We have presented an implementation of the \citetalias{G16} galaxy-targeted GW follow-up strategy using the GLADE catalog and the LCO network of telescopes. We show that the GLADE catalog is complete for galaxies brighter than $x_{1/2}$ (the median luminosity of galaxies according to a Schechter function) out to $300$\,Mpc. We detail our algorithm for selecting and prioritizing these galaxies using a GW localization and distance constraint, and taking into account the observing capabilities of LCO and the expected luminosity range of optical counterparts to NS--NS and NS--BH mergers. We discuss the results of using this strategy following two different triggers. For the second trigger, our algorithm selected the correct host galaxy as fifth in priority from the list of 182 galaxies identified in the LIGO/Virgo localization region. This allowed us to obtain early data of the counterpart \citep{Arcavi2017}. 

The galaxy prioritization algorithm presented here can be adapted to any telescope with very minor adjustments. Given the large distance to which the GLADE catalog is complete for galaxies brighter than the median galaxy luminosity, our follow-up strategy will remain relevant for following NS--NS and NS--BH GW triggers in O3.

\acknowledgments
We thank J. Canizzo for information on the \citetalias{G16} completeness plots, V. Connaughton and A. Goldstein for providing us with the Fermi GBM HEALPIX map for GRB170817A, S. Guruswamy for mentorship during S. Vasylyev's summer internship at LCO, and the LVC-EM liaisons for their guidance and assistance. Support for I.A. and J.B. was provided by NASA through the Einstein Fellowship Program, grants PF6-170148 and PF7-180162, respectively. C.M., G.H. and D.A.H. are supported by NSF grant AST-1313484. S. Vasylyev was supported by a SURF fellowship from the UCSB College of Creative Studies. D.P. and D.M. acknowledge support by ISF grant 541/17. T.P. is supported by an ERC advanced grant. D.K. is supported in part by a Department of Energy (DOE) Early Career award DE-SC0008067, a DOE Office of Nuclear Physics award DE-SC0017616, and a DOE SciDAC award DE-SC0018297, and by the Director, Office of Energy Research, Office of High Energy and Nuclear Physics, Divisions of Nuclear Physics, of the U.S. Department of Energy under Contract No.DE-AC02-05CH11231. This work made use of the LCO network. 

\begin{deluxetable*}{rDDcDDlllD}
\tablecaption{LCO follow-up observations of the Fermi GBM trigger of GRB170817A in descending order of galaxy priority. The leftmost four columns are provided as-is from the GLADE catalog. All exposures were 300 seconds long. See Table \ref{tab:lco} for the list of site abbreviations used in the telescope column.\label{tab:Fermi_followup}}
\tablehead{\colhead{GLADE} & \multicolumn2c{RA} & \multicolumn2c{Dec} & \colhead{Distance} & \multicolumn2c{m$_{\rm B}$} & \multicolumn2c{$L_{\rm B}/L_{\rm B}^*$} & \colhead{UT} & \colhead{Telescope} & \colhead{Filt.} & \multicolumn2c{Limiting Mag.} \\
\colhead{ID} & \multicolumn2c{} & \multicolumn2c{} & \colhead{[Mpc]} & \multicolumn2c{} & \multicolumn2c{} & & & & \multicolumn2c{($3\sigma$)}}
\decimals
\startdata
732352 & 204.16272 & -33.965916 & 49 & 11.24 & 4.042 & 2017-08-17 18:05:34 & CPT 1m & g & 21.88 \\
732352 & 204.16272 & -33.965916 & 49 & 11.24 & 4.042 & 2017-08-17 18:11:31 & CPT 1m & r & 21.70 \\
732352 & 204.16272 & -33.965916 & 49 & 11.24 & 4.042 & 2017-08-17 18:17:29 & CPT 1m & i & 21.10 \\
1306036 & 198.491943 & -49.478775 & 45 & 11.47 & 2.778 & 2017-08-17 18:05:34 & CPT 1m & g & 22.78 \\
1306036 & 198.491943 & -49.478775 & 45 & 11.47 & 2.778 & 2017-08-17 18:11:30 & CPT 1m & r & 22.30 \\
1306036 & 198.491943 & -49.478775 & 45 & 11.47 & 2.778 & 2017-08-17 18:17:27 & CPT 1m & i & 22.02 \\
564852 & 214.019012 & -48.127373 & 54 & 11.69 & 3.280 & 2017-08-17 18:23:50 & CPT 1m & g & 21.66 \\
564852 & 214.019012 & -48.127373 & 54 & 11.69 & 3.280 & 2017-08-17 18:29:46 & CPT 1m & r & 21.69 \\
564852 & 214.019012 & -48.127373 & 54 & 11.69 & 3.280 & 2017-08-17 23:13:25 & LSC 1m & g & 21.70 \\
564852 & 214.019012 & -48.127373 & 54 & 11.69 & 3.280 & 2017-08-17 23:19:21 & LSC 1m & r & 21.77 \\
564852 & 214.019012 & -48.127373 & 54 & 11.69 & 3.280 & 2017-08-17 23:25:16 & LSC 1m & i & 21.56 \\
2037 & 199.749649 & -47.908653 & 40 & 11.80 & 1.675 & 2017-08-17 17:47:54 & CPT 1m & g & 22.03 \\
2037 & 199.749649 & -47.908653 & 40 & 11.80 & 1.675 & 2017-08-17 17:53:51 & CPT 1m & r & 21.79 \\
2037 & 199.749649 & -47.908653 & 40 & 11.80 & 1.675 & 2017-08-17 18:23:37 & CPT 1m & g & 21.66 \\
2037 & 199.749649 & -47.908653 & 40 & 11.80 & 1.675 & 2017-08-17 18:29:33 & CPT 1m & r & 21.50 \\
2037 & 199.749649 & -47.908653 & 40 & 11.80 & 1.675 & 2017-08-17 23:13:26 & LSC 1m & g & 21.32 \\
2037 & 199.749649 & -47.908653 & 40 & 11.80 & 1.675 & 2017-08-17 23:19:22 & LSC 1m & r & 23.09 \\
2037 & 199.749649 & -47.908653 & 40 & 11.80 & 1.675 & 2017-08-17 23:25:19 & LSC 1m & i & 22.87 \\
815140 & 193.363815 & -48.749153 & 49 & 11.98 & 2.081 & 2017-08-17 17:24:22 & CPT 1m & g & 21.38 \\
815140 & 193.363815 & -48.749153 & 49 & 11.98 & 2.081 & 2017-08-17 17:30:18 & CPT 1m & r & 21.64 \\
815140 & 193.363815 & -48.749153 & 49 & 11.98 & 2.081 & 2017-08-17 17:36:15 & CPT 1m & i & 20.97 \\
1850978 & 194.305 & -46.37728 & 46 & 12.12 & 1.644 & 2017-08-17 17:24:20 & CPT 1m & g & 21.90 \\
1850978 & 194.305 & -46.37728 & 46 & 12.12 & 1.644 & 2017-08-17 17:30:17 & CPT 1m & r & 21.60 \\
1850978 & 194.305 & -46.37728 & 46 & 12.12 & 1.644 & 2017-08-17 17:36:14 & CPT 1m & i & 21.12 \\
621160 & 194.532623 & -46.264214 & 29 & 10.93 & 1.923 & 2017-08-17 17:05:17 & CPT 1m & g & 21.38 \\
621160 & 194.532623 & -46.264214 & 29 & 10.93 & 1.923 & 2017-08-17 17:11:14 & CPT 1m & r & 21.50 \\
621160 & 194.532623 & -46.264214 & 29 & 10.93 & 1.923 & 2017-08-17 17:17:10 & CPT 1m & i & 21.14 \\
737707 & 213.977814 & -48.114883 & 59 & 12.06 & 2.826 & 2017-08-17 17:05:48 & CPT 1m & g & 21.23 \\
737707 & 213.977814 & -48.114883 & 59 & 12.06 & 2.826 & 2017-08-17 17:11:47 & CPT 1m & r & 21.53 \\
737707 & 213.977814 & -48.114883 & 59 & 12.06 & 2.826 & 2017-08-17 17:17:43 & CPT 1m & i & 21.47 \\
\enddata
\end{deluxetable*}

\begin{deluxetable*}{rDDcDDlllD}
\tablecaption{LCO follow-up observations of LIGO/Virgo localization for GW170817 in descending order of galaxy priority. The leftmost four columns are provided as-is from the GLADE catalog. The galaxy which hosted the optical counterpart has GLADE ID 667146, and is fifth on this prioritized list. All exposures were 300 seconds long. See Table \ref{tab:lco} for the list of site abbreviations used in the telescope column.\label{tab:LVC_followup}}
\tablehead{\colhead{GLADE} & \multicolumn2c{RA} & \multicolumn2c{Dec} & \colhead{Distance} & \multicolumn2c{m$_{\rm B}$} & \multicolumn2c{$L_{\rm B}/L_{\rm B}^*$} & \colhead{UT} & \colhead{Telescope} & \colhead{Filt.} & \multicolumn2c{Limiting Mag.} \\
\colhead{ID} & \multicolumn2c{} & \multicolumn2c{} & \colhead{[Mpc]} & \multicolumn2c{} & \multicolumn2c{} & & & & \multicolumn2c{($3\sigma$)}}
\decimals
\startdata
1850989 & 197.465 & -24.24 & 42 & 12.78 & 0.724 & 2017-08-18 00:25:23 & LSC 1m & w & 21.89 \\
1850989 & 197.465 & -24.24 & 42 & 12.78 & 0.724 & 2017-08-19 00:02:51 & LSC 1m & w & 21.81 \\
770765 & 196.89064 & -24.008606 & 43 & 12.68 & 0.859 & 2017-08-17 23:32:42 & LSC 1m & w & 23.07 \\
770765 & 196.89064 & -24.008606 & 43 & 12.68 & 0.859 & 2017-08-18 23:55:24 & LSC 1m & w & 23.19 \\
557076 & 194.366257 & -19.691298 & 44 & 12.68 & 0.890 & 2017-08-18 09:00:23 & COJ 1m & w & 20.98 \\
557076 & 194.366257 & -19.691298 & 44 & 12.68 & 0.890 & 2017-08-18 23:47:56 & LSC 1m & w & 21.71 \\
341075 & 197.018005 & -23.796844 & 34 & 12.87 & 0.434 & 2017-08-18 00:00:22 & LSC 1m & w & 21.68 \\
341075 & 197.018005 & -23.796844 & 34 & 12.87 & 0.434 & 2017-08-19 00:02:48 & LSC 1m & w & 22.79 \\
667146 & 197.448776 & -23.383831 & 33 & 12.87 & 0.427 & 2017-08-18 00:15:23 & LSC 1m & w & 21.52 \\
667146 & 197.448776 & -23.383831 & 33 & 12.87 & 0.427 & 2017-08-18 09:10:23 & COJ 1m & w & 21.55 \\
1366038 & 197.691406 & -23.865728 & 33 & 12.64 & 0.515 & 2017-08-18 00:37:24 & LSC 1m & w & 21.70 \\
1366038 & 197.691406 & -23.865728 & 33 & 12.64 & 0.515 & 2017-08-19 00:30:02 & LSC 1m & w & 21.86 \\
1478047 & 197.466 & -24.23937 & 38 & 13.60 & 0.283 & 2017-08-17 23:40:22 & LSC 1m & w & 21.88 \\
1478047 & 197.466 & -24.23937 & 38 & 13.60 & 0.283 & 2017-08-19 00:25:22 & LSC 1m & w & 21.76 \\
602087 & 196.774902 & -23.67704 & 33 & 13.18 & 0.313 & 2017-08-18 00:07:49 & LSC 1m & w & 21.90 \\
602087 & 196.774902 & -23.67704 & 33 & 13.18 & 0.313 & 2017-08-19 00:17:50 & LSC 1m & w & 21.83 \\
773496 & 196.735474 & -23.91707 & 33 & 13.07 & 0.352 & 2017-08-18 00:15:23 & LSC 1m & w & 22.87 \\
773496 & 196.735474 & -23.91707 & 33 & 13.07 & 0.352 & 2017-08-18 23:38:53 & LSC 1m & w & 23.13 \\
645472 & 196.907242 & -23.57892 & 41 & 14.21 & 0.189 & 2017-08-17 23:50:35 & LSC 1m & w & 21.87 \\
645472 & 196.907242 & -23.57892 & 41 & 14.21 & 0.189 & 2017-08-18 23:13:50 & LSC 1m & w & 20.58 \\
3644 & 192.248566 & -14.399235 & 49 & 12.55 & 1.232 & 2017-08-18 08:39:33 & COJ 1m & w & 21.08 \\
3644 & 192.248566 & -14.399235 & 49 & 12.55 & 1.232 & 2017-08-18 23:55:24 & LSC 1m & w & 21.71 \\
626 & 193.998657 & -19.26899 & 41 & 13.27 & 0.454 & 2017-08-18 09:30:25 & COJ 1m & w & 21.84 \\
626 & 193.998657 & -19.26899 & 41 & 13.27 & 0.454 & 2017-08-19 00:10:24 & LSC 1m & w & 21.55 \\
1486718 & 196.937 & -22.85784 & 26 & 12.88 & 0.265 & 2017-08-18 00:05:02 & LSC 1m & w & 21.94 \\
1486718 & 196.937 & -22.85784 & 26 & 12.88 & 0.265 & 2017-08-18 23:40:41 & LSC 1m & w & 21.76 \\
684330 & 193.363464 & -17.005495 & 54 & 12.76 & 1.223 & 2017-08-18 09:30:25 & COJ 1m & w & 21.48 \\
684330 & 193.363464 & -17.005495 & 54 & 12.76 & 1.223 & 2017-08-18 23:21:45 & LSC 1m & w & 21.07 \\
777014 & 196.270554 & -22.383947 & 30 & 13.98 & 0.125 & 2017-08-17 23:32:43 & LSC 1m & w & 22.06 \\
777014 & 196.270554 & -22.383947 & 30 & 13.98 & 0.125 & 2017-08-18 23:46:19 & LSC 1m & w & 22.36 \\
708169 & 196.666443 & -22.455793 & 42 & 15.15 & 0.081 & 2017-08-18 08:33:18 & COJ 1m & w & 19.32 \\
708169 & 196.666443 & -22.455793 & 42 & 15.15 & 0.081 & 2017-08-19 17:15:24 & CPT 1m & w & 22.13 \\
1486724 & 197.329 & -24.38456 & 33 & 13.85 & 0.169 & 2017-08-18 00:25:23 & LSC 1m & w & 21.82 \\
1486724 & 197.329 & -24.38456 & 33 & 13.85 & 0.169 & 2017-08-19 17:25:22 & CPT 1m & w & 22.24 \\
1486713 & 196.719 & -22.84175 & 33 & 14.67 & 0.082 & 2017-08-18 08:50:25 & COJ 1m & w & 21.02 \\
1486614 & 193.706 & -16.0522 & 46 & 14.95 & 0.118 & 2017-08-18 08:40:32 & COJ 1m & w & 21.17 \\
977319 & 194.252274 & -17.320408 & 54 & 13.97 & 0.402 & 2017-08-19 17:35:24 & CPT 1m & w & 21.52 \\
1486596 & 193.107 & -15.51722 & 50 & 14.25 & 0.266 & 2017-08-18 09:00:26 & COJ 1m & w & 20.78 \\
1490961 & 197.177 & -23.77574 & 39 & 15.22 & 0.066 & 2017-08-18 18:20:21 & CPT 1m & w & 21.21 \\
1490961 & 197.177 & -23.77574 & 39 & 15.22 & 0.066 & 2017-08-19 17:06:27 & CPT 1m & w & 21.68 \\
420937 & 198.880432 & -23.982388 & 35 & 12.32 & 0.776 & 2017-08-18 09:57:07 & COJ 1m & w & 21.64 \\
635635 & 196.600052 & -24.164007 & 33 & 13.78 & 0.180 & 2017-08-18 09:46:43 & COJ 1m & w & 21.68 \\
635635 & 196.600052 & -24.164007 & 33 & 13.78 & 0.180 & 2017-08-19 17:15:22 & CPT 1m & w & 21.67 \\
7 & 192.519547 & -14.73349 & 53 & 13.21 & 0.782 & 2017-08-18 09:21:21 & COJ 1m & w & 21.59 \\
7 & 192.519547 & -14.73349 & 53 & 13.21 & 0.782 & 2017-08-19 17:35:25 & CPT 1m & w & 21.90 \\
795473 & 199.096786 & -26.561554 & 44 & 13.95 & 0.271 & 2017-08-18 09:55:25 & COJ 1m & w & 22.04 \\
645300 & 196.580811 & -22.98033 & 39 & 15.71 & 0.042 & 2017-08-18 09:21:22 & COJ 1m & w & 21.63 \\
645300 & 196.580811 & -22.98033 & 39 & 15.71 & 0.042 & 2017-08-19 17:25:23 & CPT 1m & w & 21.75 \\
1486721 & 197.135 & -23.34725 & 44 & 15.71 & 0.054 & 2017-08-18 08:50:23 & COJ 1m & w & 20.93 \\
646603 & 193.219254 & -15.413292 & 56 & 13.09 & 0.961 & 2017-08-18 09:10:23 & COJ 1m & w & 21.50 \\
646603 & 193.219254 & -15.413292 & 56 & 13.09 & 0.961 & 2017-08-19 17:06:22 & CPT 1m & w & 21.24 \\
1566 & 192.827362 & -14.573568 & 51 & 13.80 & 0.427 & 2017-08-18 17:15:25 & CPT 1m & w & 21.25 \\
3385 & 199.958435 & -27.410082 & 27 & 10.97 & 1.644 & 2017-08-18 17:42:31 & CPT 1m & w & 23.25 \\
3385 & 199.958435 & -27.410082 & 27 & 10.97 & 1.644 & 2017-08-19 09:57:51 & COJ 1m & w & 22.84 \\
1308288 & 196.907013 & -23.938364 & 45 & 15.59 & 0.064 & 2017-08-18 18:15:26 & CPT 1m & w & 21.95 \\
1308288 & 196.907013 & -23.938364 & 45 & 15.59 & 0.064 & 2017-08-19 09:35:25 & COJ 1m & w & 22.57 \\
3250 & 199.521057 & -26.837221 & 24 & 11.10 & 1.117 & 2017-08-18 18:30:25 & CPT 1m & w & 21.30 \\
3250 & 199.521057 & -26.837221 & 24 & 11.10 & 1.117 & 2017-08-19 09:17:51 & COJ 1m & w & 21.56 \\
2704 & 193.830368 & -14.949816 & 43 & 14.31 & 0.185 & 2017-08-18 17:15:25 & CPT 1m & w & 21.76 \\
2704 & 193.830368 & -14.949816 & 43 & 14.31 & 0.185 & 2017-08-19 08:45:22 & COJ 1m & w & 21.19 \\
1866 & 194.690292 & -17.542887 & 54 & 14.87 & 0.175 & 2017-08-18 18:05:41 & CPT 1m & w & 21.81 \\
1866 & 194.690292 & -17.542887 & 54 & 14.87 & 0.175 & 2017-08-19 09:25:34 & COJ 1m & w & 21.38 \\
4242 & 189.997894 & -11.62307 & 9 & 8.52 & 1.959 & 2017-08-19 09:10:20 & COJ 0.4m & w & 19.75 \\
1486716 & 196.782 & -24.11136 & 27 & 14.07 & 0.090 & 2017-08-18 18:10:25 & CPT 1m & w & 21.58 \\
1486716 & 196.782 & -24.11136 & 27 & 14.07 & 0.090 & 2017-08-19 09:41:24 & COJ 1m & w & 21.86 \\
665505 & 193.212875 & -15.491673 & 50 & 15.38 & 0.095 & 2017-08-18 17:25:23 & CPT 1m & w & 21.95 \\
665505 & 193.212875 & -15.491673 & 50 & 15.38 & 0.095 & 2017-08-19 09:30:04 & COJ 1m & w & 21.75 \\
1220861 & 192.654388 & -14.482746 & 49 & 15.26 & 0.100 & 2017-08-18 17:25:21 & CPT 1m & w & 21.52 \\
1220861 & 192.654388 & -14.482746 & 49 & 15.26 & 0.100 & 2017-08-19 08:33:51 & COJ 1m & w & 20.17 \\
761543 & 197.599045 & -21.684093 & 35 & 13.85 & 0.190 & 2017-08-18 17:35:04 & CPT 1m & w & 22.19 \\
761543 & 197.599045 & -21.684093 & 35 & 13.85 & 0.190 & 2017-08-19 09:50:23 & COJ 1m & w & 21.74 \\
720029 & 193.620285 & -16.350813 & 55 & 14.90 & 0.179 & 2017-08-18 17:05:47 & CPT 1m & w & 20.65 \\
720029 & 193.620285 & -16.350813 & 55 & 14.90 & 0.179 & 2017-08-19 09:00:38 & COJ 1m & w & 21.83 \\
1071538 & 192.717911 & -14.906625 & 50 & 15.28 & 0.106 & 2017-08-18 18:00:19 & CPT 1m & w & 21.39 \\
1071538 & 192.717911 & -14.906625 & 50 & 15.28 & 0.106 & 2017-08-19 09:10:25 & COJ 1m & w & 21.36 \\
2151 & 191.728119 & -11.637039 & 29 & 13.26 & 0.233 & 2017-08-18 17:52:16 & CPT 1m & w & 21.49 \\
2151 & 191.728119 & -11.637039 & 29 & 13.26 & 0.233 & 2017-08-19 08:55:23 & COJ 1m & w & 21.05 \\
1486720 & 197.064 & -21.00158 & 34 & 15.19 & 0.053 & 2017-08-18 17:06:09 & CPT 1m & w & 21.19 \\
1486720 & 197.064 & -21.00158 & 34 & 15.19 & 0.053 & 2017-08-19 09:42:50 & COJ 1m & w & 21.38 \\
1481025 & 197.324 & -24.38207 & 38 & 16.50 & 0.019 & 2017-08-18 18:25:24 & CPT 1m & w & 21.85 \\
1481025 & 197.324 & -24.38207 & 38 & 16.50 & 0.019 & 2017-08-19 09:02:53 & COJ 1m & w & 21.59 \\
1486644 & 194.643 & -16.80437 & 52 & 14.81 & 0.175 & 2017-08-19 08:55:20 & COJ 0.4m & w & 19.99 \\
811204 & 201.209473 & -30.307772 & 53 & 12.70 & 1.240 & 2017-08-19 09:57:15 & COJ 0.4m & w & 20.46 \\
1490974 & 196.879 & -23.17047 & 40 & 17.26 & 0.011 & 2017-08-19 08:40:19 & COJ 0.4m & w & 19.75 \\
1490971 & 196.348 & -23.52258 & 39 & 16.10 & 0.030 & 2017-08-19 08:50:11 & COJ 0.4m & w & 19.99 \\
722418 & 198.573929 & -26.58268 & 53 & 14.63 & 0.213 & 2017-08-19 09:35:20 & COJ 0.4m & w & 20.44 \\
519820 & 201.013931 & -32.341335 & 35 & 12.45 & 0.683 & 2017-08-19 09:02:15 & COJ 0.4m & w & 20.67 \\
336095 & 201.988586 & -31.499374 & 35 & 12.13 & 0.908 & 2017-08-19 10:00:14 & COJ 0.4m & w & 19.86 \\
607497 & 199.752182 & -27.628489 & 29 & 13.20 & 0.233 & 2017-08-19 09:50:10 & COJ 0.4m & w & 20.68 \\
640513 & 192.717209 & -14.491902 & 57 & 13.58 & 0.644 & 2017-08-19 09:10:11 & COJ 0.4m & w & 19.79 \\
1486585 & 192.723 & -14.33199 & 57 & 13.41 & 0.756 & 2017-08-19 09:00:29 & COJ 0.4m & w & 19.95 \\
1486633 & 194.261 & -19.51809 & 56 & 14.48 & 0.273 & 2017-08-19 09:25:20 & COJ 0.4m & w & 19.72 \\
751761 & 197.648849 & -21.748224 & 28 & 13.33 & 0.198 & 2017-08-19 08:45:22 & COJ 0.4m & w & 20.00 \\
1478083 & 201.121 & -30.43168 & 52 & 13.45 & 0.614 & 2017-08-19 09:17:16 & COJ 0.4m & w & 21.75 \\
592826 & 199.083679 & -28.285717 & 59 & 13.07 & 1.094 & 2017-08-19 09:42:13 & COJ 0.4m & w & 20.50 \\
341078 & 194.386322 & -19.700184 & 50 & 16.18 & 0.045 & 2017-08-19 08:33:32 & COJ 0.4m & w & 20.57 \\
\enddata
\end{deluxetable*}

\bibliography{refs}

\end{document}